\documentclass[12pt]{amsart}
 \usepackage{amsmath,amsthm}
 \usepackage{amsfonts,amscd,amssymb,latexsym}
 \usepackage[matrix,arrow,curve]{xy}
 \usepackage{rotating,array}
 \usepackage[english]{babel}
 \usepackage{graphicx}

 \theoremstyle{plain}
 \newtheorem{theorem}{Theorem}

 \theoremstyle{definition}

 \theoremstyle{remark}

 \makeindex

\begin{document}

 \title[GINZBURG--LANDAU EQUATIONS]{GINZBURG--LANDAU EQUATIONS AND THEIR GENERALLIZATIONS}
 \author{Armen SERGEEV}

\dedicatory{Dedicated to 100th anniversary of Professor Jaap KOREVAAR}

 \maketitle


\section*{Introduction}
\label{sec0}


The Ginzburg--Landau equations were proposed in the superconductivity
theory to describe mathematically the intermediate state of superconductors
in which the normal conductivity is mixed with the superconductivity
(cf. \cite{LP}). It was understood later on that these equations
play an important role also in various problems of mathematical physics.
We mention here the extension of these equations to compact Riemann surfaces
and Riemannian 4-manifolds. A separate interesting topic is the scattering
theory of vortices reducing to the study of hyperbolic Ginzburg--Landau equations.
In this review we tried to touch these interesting topics with many
still unsolved problems.

Briefly on the content of the paper. We start from Section \ref{sec1} in which we introduce
the Ginzburg--Landau equations on the plane. The physical aspects of these
equations are described in \cite{LP} (cf. also \cite{Ser}).
In Subsection \ref{ssec1.2} we describe the vortex solutions which are the local minima
of the functional of potential energy. The main result here is the description
of the $d$-vortex solutions due to Taubes (cf. \cite{JT}). The Subsection \ref{ssec1.3}
is devoted to the generalization of the results of previous two subsections
to compact Riemann surfaces. The generalization of Taubes theorem to
this case was obtained by Bradlow in \cite{B}.

In Section \ref{sec2} we switch on the time variable and consider the hyperbolic
Ginzburg--Landau equations introduced in Subsection \ref{ssec2.1}.
We study the adiabatic or slow time limit in these equations.
The hyperbolic Ginzburg--Landau equations in this limit convert into the
adiabatic equations. Their solutions, called the
adiabatic trajectories, are given by the geodesics on the
moduli space of vortex solutions with respect to the metric
generated by the kinetic energy functional. Solving the Euler
equation for these geodesics, we can describe approximately
solutions of the original Ginzburg--Landau equations with small
kinetic energy.

In Section \ref{sec3} we deal with the Seiberg--Witten
equations which may be considered as an extension of Ginzburg--Landau equations
to compact Riemannian 4-manifolds. A key idea is to use the
$\text{Spin}^c$-structure existing on any Riemannian 4-manifold.
Necessary notions from the spinor geometry are given in Subsection 3.1.
In Subsection 3.2 we introduce the Seiberg--Witten equations on
compact Riemannian 4-manifolds. In the next Subsection 3.3
we consider the model example of
Seiberg--Witten equations on a compact K\"ahler surface.
In this case the moduli space of solutions
coincides with the space of holomorphic curves on the
considered surface lying in a given topological class.
In the last Subsection 3.4 we study the
Seiberg--Witten equations on compact symplectic 4-manifolds.
Again, as in the 3-dimensional case, we use the adiabatic limit construction
with scale parameter $\lambda\to\infty$. In the limit the
sequence of solutions of the Seiberg--Witten equations, depending on the
scale parameter, converges (in the weak sense) to a pseudoholomorphic curve
which may be considered as a complex analogue of the adiabatic trajectory
in the 3-dimensional case.
The parameter along this limiting curve plays the role of the
''complex time''. The Seiberg--Witten equations in this
limit reduce to a family of vortex equations, defined in the normal
planes to the limiting pseudoholomorphic curve. The limiting curve
and a family of vortex solutions along this curve must satisfy the
adiabatic equation analogous to the $\bar\partial$-equation
which may be considered as a complex analogue of the Euler geodesic equation.
Conversely, if we have a pseudoholomorphic curve and a family of
vortex equations in normal planes, satisfying the adiabatic
equation, then we can reconstruct from these data a solution of
Seiberg--Witten equations which tends in the adiabatic limit to the
original pseudoholomorphic curve and given family of vortex
solutions.

While preparing this paper the author was partially supported financially
by the RSF grant 19-11-00316.

\section{Ginzburg--Landau equations in dimension 2}
\label{sec1}

\subsection{Ginzburg--Landau Lagrangian}
\label{ssec1.1}

The Ginzburg--Landau equations on the plane $\mathbb R^2_{(x_1,x_2)}$
are the Euler--Lagrange equations for the potential energy functional
of the form
\begin{equation}
\label{eq11}
U(A,\Phi):=\frac{1}{2}\int\mathcal{L}(A,\Phi)\,dx_1dx_2
\end{equation}
where $\mathcal{L}(A,\Phi)$ is the Ginzburg--Landau Lagrangian. This
Lagrangian depends on two variables $A$ and $\Phi$. The first of them is
the 1-form
$$
A=A_1dx_1+A_2dx_2
$$
with smooth pure imaginary coefficients. The second one is a smooth
complex-valued function $\Phi=\Phi_1+i\Phi_2$.

The Lagrangian $\mathcal{L}(A,\Phi)$ has the form
\begin{equation}
\label{eq12}
\mathcal{L}(A,\Phi)=|F_A|^2+|D_A\Phi|^2+\frac14\,(1-|\Phi|^2)^2.
\end{equation}
Here
$$
F_A=dA=\sum_{i,j=1}^2F_{ij}dx_i\wedge dx_j=2F_{12}dx_1\wedge dx_2
$$
where $F_{ij}=\partial_iA_j-\partial_jA_i$ with $\partial_i:=
\partial/\partial x_{i}$,
$$
d_A\Phi=d\Phi+A\Phi=\sum_{i=1}^2(\partial_i+A_i)\Phi\,dx_i.
$$

The Euler--Lagrange equations for the potential energy functional
\eqref{eq11} have the form
\begin{equation}
\label{eq13}
\left\{
\begin{aligned}
\partial_i F_{ij} &=0, \quad j=1,2,\\
\nabla^{2}_{A} \Phi &=\frac12\,\Phi(|\Phi|^2-1).
\end{aligned}\right.
\end{equation}

To satisfy the condition $U(A,\Phi)<\infty$ we shall require that
$|\Phi| \to 1$ for $|x|\to\infty$. It follows from this asymptotic
condition that our problem has a topological invariant given by
the rotation number $d$ of the map $\Phi$ sending the circles $S^{1}_{R}$
of large radius $R$ to topological circles $|\Phi|\approx 1$.
This invariant takes on the integer values and is called the vortex number.

\subsection{Vortex equations}
\label{ssec1.2}

Assume now that $d\geq0$. It may be proved that
$$
U(A,\Phi)\geq\pi d
$$
and the equality here is attained only on solutions of the vortex equations
which are written in complex coordinate $z=x_1+ix_2$ in the form
\begin{equation}
\label{vortex}
\left\{
\begin{aligned}
\bar\partial_A\Phi &=0,\\
iF_{12} &=\frac12(1-|\Phi |^2)
\end{aligned}\right.
\end{equation}
where $\bar\partial_A\Phi=\bar\partial+A^{0,1}$ with $A^{0,1}$ being the
$(0,1)$-component of the form $A$ written in terms of complex coordinate like
$A=A^{1,0}+A^{0,1}$.

Note that the vortex equations, as well as potential energy $U(A,\Phi)$, are
invariant under gauge transforms given by
$$
A\longmapsto A+id\chi,\quad \Phi\longmapsto e^{-i\chi} \Phi
$$
where $\chi$ is a smooth real-valued function.

For $d<0$ we have the similar inequality
$$
U(A,\Phi)\geq-\pi d
$$
where the equality is attained on solutions of anti-vortex equations
$$
\left\{
\begin{aligned}
\partial_A \Phi &=0,\\
iF_{12} &=\frac12\,(|\Phi|^2-1).
\end{aligned}\right.
$$

Solutions of the vortex equations are described by the following theorem.

\begin{theorem}[Taubes \cite{JT}]
\label{T1}
For any natural number $d>0$ and arbitrary collection
$\{z_1,z_2,\ldots z_k\}$ of different points in the complex plane
$\mathbb C$, taken with multiplicities $d_1,d_2,\ldots d_k$ such
that $\sum_{j=1}^kd_j=d$, there exists a unique (up to gauge
transforms) $d$-vortex solution $(A,\Phi)$ with $U(A,\Phi)<\infty$,
satisfying the condition: the divisor of zeros of the function
$\Phi$ coincides with $\sum_{j=1}^k d_jz_j$.
\end{theorem}

An analogous theorem holds for solutions of anti-vortex equations for
$d<0$. Moreover, Taubes has proved that any critical point $(A,\Phi)$ of
the potential energy \eqref{eq11} with $U(A,\Phi)<\infty$ and $d>0$ is gauge
equivalent to some $d$-vortex solution.
It follows that any solution of the Euler--Lagrange equations with $U(A,\Phi)<\infty$
is either $d$-vortex, or $|d|$-anti-vortex.

The moduli space of $d$-vortices is by definition the quotient
$$
\mathcal M_d=\frac{\{d\text{-vortices}\ (A,\Phi)} {\{\text{gauge
transforms}\}}\,.
$$
In the sequel we restrict to the case $d>0$.

Theorem \ref{T1} implies that the moduli space of $d$-vortices
coincides with the set of unordered collections of $d$ points in the
complex plane $\mathbb C$, i.e. with the $d$th \textit{symmetric
power} of $\mathbb C$:
$$
\mathcal M_d=\text{Sym}^d\mathbb C.
$$
Note that the symmetric power $\text{Sym}^d\mathbb C$ may be
identified with the space $\mathbb C^d$ by assigning to the
collection of $d$ points in the complex plane $\mathbb C$ the
polynomial with the highest coefficient equal to 1, having its zeros at
given points.

\subsection{Vortex equations on compact Riemann surfaces}
\label{ssec1.3}

Let $X$ be a compact Riemann surface provided with Riemannian metric
$g$ and K\"ahler form $\omega$. We fix a complex Hermitian line
bundle $L\to X$ with Hermitian metric $h$ and define the energy
functional by analogy with the complex plane case
$$
U(A,\Phi)=\frac12\int_X \left\{|F_A|^2+|d_A \Phi|^2+
\frac14(1-|\Phi|^2)^2\right\}\omega.
$$
Here, $A$ is a $\text{U}(1)$-connection on $L$, $F_A= dA$ is its
curvature, $d_A$ is the covariant exterior derivative, generated by
$A$, $\Phi$ is a section of the bundle $L\to X$, its norm $|\Phi|$
being computed with respect to metric $h$. As in the complex plane
case, this functional is invariant under gauge transforms, given by
the maps $u\in\mathcal G= C^\infty(X,U(1))$.

The first Chern class $c_1(L)$ of the line bundle $L\to X$ is equal,
according to Gauss-Bonnet formula, to
$$
c_1(L)=\frac{i}{2\pi}\int_X F_A.
$$

Let us assume that $c_1 (L)>0$. Then, as in Subsection \ref{ssec1.2}, we have the lower estimate
for the energy of the form
$$
U(A,\Phi)\geq\pi c_1(L).
$$
The equality here is attained on solutions of the equations
\begin{equation}
\label{compact}
\left\{
\begin{aligned}
\bar\partial_A\Phi &=0,\\
iF_{A}^{\omega} &=\frac12(1-|\Phi|^2)
\end{aligned}\right.
\end{equation}
where $F_{A}^{\omega}=\omega\lrcorner\, F_A$ is the (1,1)-component of
the curvature $F_A$, parallel to $\omega$.

The obtained equations look the same as the vortex equations on the
complex plane. However, in the case of a compact Riemann surface we
have an evident obstruction to their solvability. Namely, by
integrating the second equation over $X$, we get
$$
\frac{i}{2\pi}\int_XF_A=\frac1{4\pi}\int_X\omega-\frac1{4\pi}\int_X|\Phi|^2\omega,
$$
which may be rewritten in the form
$$
c_1(L)=\frac1{4\pi}\text{Vol}_g(X)-\frac1{4\pi}\Vert \Phi\Vert^{2}_{L^2}.
$$
So we arrive at the \textit{necessary condition of the solvability}
of the above equations:
$$
c_1(L)\leq\frac1{4\pi}\text{Vol}_g(X).
$$
This condition arises because of the
non-invariance of the energy under the scale transform.

The scale transform changes the metric $g$ to
the metric $g_t:=t^2g$. Simultaneously, the K\"ahler form and volume
change to:
$$
\omega_t= t^2\omega, \quad\text{Vol}_{g_t}(X)=t^2\text{Vol}_{g}(X).
$$
The necessary solvability condition for the rescaled metric $g_t$
looks like:
$$
c_1(L)\leq\frac{t^2}{4\pi}\text{Vol}_g (X)
$$
and is evidently satisfied for sufficiently large
$t$. So we can always attain the necessary
solvability condition by rescaling the
original metric $g$.

It is, however, more convenient to fix the metric and introduce the
scaling into the definition of the functional $U(A,\Phi)$. Namely,
we replace the energy functional $U(A,\Phi)$ by its rescaled
version
$$
U_{\tau}(A,\Phi)= \frac12\int_X\left\{|F_A|^2+|d_A
\Phi|^2+\frac12(\tau-|\Phi|^2)^2\right\}
$$
where $\tau >0$ is the scaling parameter.

Then we obtain the following lower estimate for the energy
$$
U_{\tau}(A,\Phi)\geq\pi c_1(L),
$$
where the equality is attained only on solutions of the equations
\begin{equation}
\label{righteq}
\left\{
\begin{aligned}
\bar\partial_A\Phi &=0,\\
iF_{A}^{\omega} &=\frac12(\tau-|\Phi|^2).
\end{aligned}\right.
\end{equation}
These are the right vortex equations on a compact Riemann
surface. For them the
necessary solvability condition takes the form
$$
c_1(L)\leq\frac{\tau}{4\pi}\text{Vol}_g (X).
$$

For these equations we have the following analogue of the
Taubes theorem.

\begin{theorem}[Bradlow \cite{B}]
\label{bradlow} Let $d:=c_1(L)>0$ and $D$ is an effective divisor on
$X$ of degree $d$, i.e. $D=\sum_{j=1}^kd_jz_j$ with
$\sum_{j=1}^kd_j=d $. Then the condition
$$
c_1(L)<\frac{\tau}{4\pi}\mathrm{Vol}(X)
$$
is necessary and sufficient for the existence of a unique (up to
gauge equivalence) $d$-vortex solution $(A,\Phi)$ such that the zero
divisor of $\Phi$ coincides with $D$.

Moreover, the holomorphic line bundle $L$, provided with the complex
structure determined by the operator $\bar\partial_A$, is isomorphic
to the holomorphic line bundle $[D]$, defined by the divisor $D$.
\end{theorem}

Note that the first vortex equation $\bar\partial_A\Phi =0$ means,
in other words, that $\Phi$ is a holomorphic section of the
Hermitian line bundle $(L,\bar\partial_A)$.

According to Bradlow theorem, in the case when
$$
c_1(L)<\frac{\tau}{4\pi}\,\text{Vol}(X)
$$
we have a bijective correspondence between the sets:
$$
\{d\text{-vortex solutions}\ (A,\Phi)\}/\mathcal{G}
$$
and
$$
\{\text{\,effective divisors}\ D\ \text{of degree}\, d=c_1(L)\}.
$$
So the moduli space of $d$-vortex solutions coincides with the
symmetric power $\text{Sym}^d X$.

The inequality
$$
\tau>\frac{4\pi c_1(L)}{\text{Vol}(X)}
$$
coincides with the stability condition for the pair
$(E,\Phi)$ (cf. \cite{B}).

\section{Ginzburg--Landau equations in dimension 3}
\label{sec2}

\subsection{Hyperbolic Ginzburg--Landau equations}
\label{ssec2.1}

We add the time variable $x_0=t$ to the variables $(x_1,x_2)$
and denote by $\Phi=\Phi(t,x_1,x_2)$
a smooth complex-valued function on the space $\mathbb R^3=\mathbb R^{1+2}$ with
coordinates ${(t,x_1,x_2)}$. The form $A$ from Section \ref{sec1} is replaced by the
form
$$
\mathcal A=A_0dt+A_1dx_1+A_2dx_2
$$
with coefficients $A_\mu=A_\mu(t,x_1,x_2)$, $\mu=0,1,2$, being
smooth functions with pure imaginary values on the space $\mathbb
R^{1+2}$. Denote the time component of the form $\mathcal A$ by
$A^0:=A_0dt$ and its space component by
$A=A_1dx_1+A_2dx_2$.

The potential energy of the system is given by
the same formula, as in Subsection \ref{ssec1.1}, i.e. $U(\mathcal
A,\Phi)=U(A,\Phi)$.

We define the kinetic energy of the system by
$$
T(\mathcal A,\Phi)=\frac12\int\left\{|F_{01}|^2+|F_{02}|^2+
|d_{A^0}\Phi|^2\right\}\,dx_1dx_2
$$
where $F_{0j}$, $j=1,2$, are given, as before, by the formula
$$
F_{0j}=\partial_0 A_j-\partial_j A_0,
$$
and $d_{A^0}\Phi=d\Phi+A_0\,dt$.

Introduce the Ginzburg--Landau action functional
$$
S(\mathcal A,\Phi) = \int_0^{T_0}\left(T(\mathcal A,\Phi) -
U(\mathcal A,\Phi)\right)\,dt.
$$
The Euler--Lagrange equations for this functional, called also the
hyperbolic Ginzburg--Landau equations, have the form:
$$
\left\{\begin{aligned}
\partial_1F_{01} &+\partial_2F_{02}= i\,\text{Im}(\bar\Phi\nabla_{A,0}\Phi)\\
\partial_0F_{0j} &+ \sum_{k=1}^2\varepsilon_{jk}\partial_kF_{12}=
i\,\text{Im}(\bar\Phi\nabla_{A,j}\Phi),\ j=1,2 \\
(\nabla_{A,0}^2 &-\nabla_{A,1}^2-\nabla_{A,2}^2)\Phi=
\frac12\Phi(1-|\Phi|^2)
\end{aligned}\right.
$$
where $\nabla_{A,\mu}=\partial_{\mu}+A_{\mu},\ \mu=0,1,2$,
$\varepsilon_{12}=-\varepsilon_{21}=1$,
$\varepsilon_{11}\varepsilon_{22}=0$.

The first of these equations is of constraint type which
means that it holds for any $t$ if it is satisfied for the initial
moment of time. The last equation, containing the
covariant D'Alembertian in its left hand side, is a
nonlinear wave equation.

These equations are invariant under the gauge transforms of
the form
$$
A\mapsto A+id\chi,\quad\Phi\mapsto e^{-i\chi}\Phi
$$
where $\chi=\chi(t,x_1,x_2)$ is a smooth real-valued function on
$\mathbb R^{1+2}$.

We can choose the gauge function $\chi$ so that $A_0=0$, such a
choice is called the temporal gauge.
(Note that after fixing the temporal gauge we still have the gauge
freedom with respect to static gauge transforms, given by gauge
functions $\chi$ which do not depend on time $t$.)

In the temporal gauge the kinetic energy is written in the form
$$
T(A,\Phi)=\frac12\{\Vert\dot{\Phi}\Vert^2+\Vert\dot{A}\Vert^2\}
$$
where ''dot'' denotes the time derivative $\partial/\partial
t=\partial/\partial x_0$ and
$\Vert\cdot\Vert=\Vert\cdot\Vert_{L^2(\mathbb R^2)}$ is the norm in
the space $L^2(\mathbb R^2)$.

\subsection{Adiabatic limit}
\label{ssec2.2}

Our goal is to describe the space of solutions of hyperbolic
Ginzburg--Landau equations modulo dynamic gauge transforms. We shall
call the solutions of these equations, for brevity, the
dynamic solutions and the quotient of the space of dynamic
solutions modulo gauge transforms is called the moduli
space of dynamic solutions.

In contrast with the case of the moduli space of static solutions,
which structure is completely described by Taubes theorems, we
cannot expect to get anything similar in the dynamic case. However,
we can hope to obtain an approximate description of at least some
classes of dynamic solutions. We shall present here
an heuristic approach, proposed by Manton (cf. \cite{Ma}), to
the approximate description of ''slowly moving'' dynamic solutions.

In the temporal gauge the dynamic solutions of hyperbolic
Ginzburg--Landau equations are given by the smooth trajectories
$$
\gamma: t\longmapsto[A(t),\Phi(t)]
$$
in the static configuration space:
$$
\mathcal N_d=\frac{\{\text{smooth data}\ (A,\Phi)\ \text{with}\
U(A,\Phi)<\infty\ \text{and vortex number}\ d\}} {\{\text{static
gauge transforms}\}}
$$
where $[A(t),\Phi(t)]$ denotes the gauge class of the pair
$(A(t),\Phi(t))$ modulo static gauge transforms. It contains, in
particular, the moduli space $\mathcal M_d$ of $d$-vortex solutions.

The configuration space $\mathcal N_d$ may be thought of as a
horizontal canyon with a small ball with trajectory $\gamma(t)$,
rolling inside it. The moduli space of $d$-vortex solutions
$\mathcal M_d$, for which the potential energy is minimal,
corresponds to the bottom of this canyon. The lower is the kinetic
energy of the ball, the closer lies its trajectory to the bottom of
the canyon. The ball can even hit this bottom but, having a non-zero
kinetic energy, cannot stop there and is forced to assent the wall
of the canyon.

Define the kinetic energy of the trajectory $\gamma(t)=[A(t),\Phi(t)]$ by
$$
T(\gamma):=\frac12\{\Vert\dot{A}\Vert^2+\Vert\dot{\Phi}\Vert^2\}.
$$

Consider the family of trajectories $\gamma_\varepsilon(t)$,
depending on the parameter $\varepsilon>0$, having the kinetic
energy $T(\gamma_\varepsilon)$ proportional to
$\varepsilon$. For
small $\varepsilon$ the trajectories $\gamma_\varepsilon(t)$ are
lying close to the moduli space $\mathcal M_d$ and
in the limit $\varepsilon\to 0$ they converge to a static
solution, i.e. to a point on $\mathcal M_d$.

However, if we introduce the ''slow time''
$\tau:=\varepsilon t$ on the trajectory $\gamma_\varepsilon$
then in the limit $\varepsilon \to 0$ the ''rescaled''
trajectories $\gamma_\varepsilon(\tau)$ will converge not to a point
but to some trajectory $\gamma_0$, lying in $\mathcal M_d$.

The described construction is called the adiabatic limit and the equations, to which
the original Ginzburg--Landau equations reduce in this limit, are called
the adiabatic equations. Accordingly, their solutions are called the
adiabatic trajectories.

The adiabatic trajectories admit the following intrinsic description
in terms of the moduli space $\mathcal M_d$.

\begin{theorem}
\label{adi_traj} The kinetic energy functional determines a
Riemannian metric on the vortex space $\mathcal M_d$, called the
kinetic or T-metric. Geodesics of this metric
coincide precisely with the adiabatic trajectories.
\end{theorem}

Since any point of an adiabatic trajectory $\gamma_0$ is a static
solution, the trajectory itself cannot be a dynamic solution.
However, such trajectories describe approximately dynamic solutions
with small kinetic energy.

Manton has formulated the following adiabatic principle:
\textit{For any adiabatic trajectory $\gamma_0$ on the moduli space
$\mathcal M_d$ it should exist a sequence $\{\gamma_\varepsilon\}$
of dynamic trajectories (solutions of hyperbolic Ginzburg--Landau
equations), converging for $\varepsilon\to 0$ to $\gamma_0$ in the
adiabatic limit.}

The rigorous formulation of this principle and its proof were given
by Roman Palvelev \cite{Pal} (cf.also \cite{Pal-Ser}).

\section{Seiberg--Witten equations}
\label{sec3}

In this section we consider one more generalization of Ginzburg--Landau
equations, this time to 4-dimensional Riemannian manifolds. These are
the Seiberg--Witten equations. Let us start with some basic definitions
from spinor geometry of 4-manifolds.

\subsection{Spinor geometry}
\label{ssec3.1}

A key role in studying the 4-dimensional Riemannian manifolds is
played by the $\text{Spin}^c$-structure which exists on any
Riemannian 4-manifold. It can be considered as a replacement of the
complex structure underlying the theory of 2-dimensional Riemann
surfaces.

Leaving apart the general definition of $\text{Spin}^c$-structure (which may be found
in the book \cite{LM}) we describe here its properties
used in Seiberg--Witten theory.

Let $(X,g)$ be a compact oriented Riemannian 4-manifold provided
with Levi-Civita connection. Then we can define the Clifford
multiplication $\rho$ by differential forms on $X$, i.e. a representation of
such forms by linear endomorphisms acting on smooth sections of the
spinor bundle $W\to X$. It is a complex Hermitian vector
bundle of rank 4 decomposed into the direct sum
$$
W=W^+\oplus W^-
$$
of complex semispinor bundles of rank 2.

The spinor bundle $W$ may be provided with spinor
connection $\nabla$ which is an extension of the Levi-Civita
connection to a connection on $W$. The Dirac
operator on smooth sections of $W$ is given by the composition
$\rho\circ\nabla$ of Clifford multiplication with spinor connection.

In the case when the manifold $(X,g)$ is
symplectic, i.e. provided with the symplectic form
$\omega$ compatible with $g$, it also has an almost
complex structure $J$ compatible both with $\omega$ and $g$.

In this case we have a canonical construction of the spinor bundle
$W$ identified with
$$
W_{\text{can}}=\Lambda^{0,*}(T^*X)=\bigoplus_{q=0}^{2}\Lambda^{0,q}(T^*X).
$$
Accordingly,
$$
W_{\text{can}}^+=\Lambda^{0,0}(T^*X)\oplus\Lambda^{0,2}(T^*X)\,,\quad
W_{\text{can}}^-=\Lambda^{0,1}(T^*X).
$$
In this case there is also a canonical spinor connection
$\nabla_{\text{can}}$ on $W_{\text{can}}$ and an explicit formula for the Clifford
multiplication (cf. \cite{LM},\cite{Ser}).

Moreover, for any Hermitian line bundle $E\to X$ with a Hermitian
connection $B$ on it we can construct the associated spinor bundle
$W_E:=W_{\text{can}}\otimes E$ and the spinor connection $\nabla_A$ on
$W_E$ where $A=A_E$ is the tensor product of the canonical spinor
connection $\nabla_{\text{can}}$ on $W_{\text{can}}$ and the given Hermitian
connection $B$ on $E$.

The Dirac operator
$$
D_A=\rho\circ\nabla_A:\Gamma(X,W^+)\longrightarrow\Gamma(X,W^-)
$$
coincides in this case with
$\bar\partial_B+\bar\partial_B^*$ where $\bar\partial_B^*$ is the
$L^2$-adjoint of the operator $\bar\partial_B$.

\subsection{Seiberg--Witten equations on Riemannian 4-manifolds}
\label{ssec3.2}

Let $(X,g)$ be a compact oriented Riemannian 4-manifold provided
with a $\text{Spin}^c$-structure and $E\to X$ is Hermitian line bundle
provided with a Hermitian connection $B$.

Consider the following Seiberg--Witten action functional
$$
S(A,\Phi)=\frac12\int_X\left\{|F_A|^2+|\nabla_A\Phi|^2+
\left(\text{s}(g)+|\Phi|^2\right)\frac{|\Phi|^2}{4}\right\}\text{vol}
$$
where $\text{s}(g)$ is the scalar curvature of $(X,g)$, $F_A$ is
the curvature of the connection $\nabla_A$, $\Phi$ is a smooth section of $W^+$ and $\text{vol}$
is the volume element on $(X,g)$.

The local minima of this functional satisfy the
Seiberg--Witten equations
\begin{equation}
\left\{
\begin{aligned}
\label{eq1}
D_{A}\Phi &=0,\hspace{2cm}\\
F_{A}^{+}=&\ \Phi\otimes\Phi^{*}-\frac12|\Phi|^2\cdot\text{Id}
\end{aligned}\right.
\end{equation}
where $\Phi\otimes\Phi^{*}-\frac12|\Phi|^2\cdot\text{Id}$ is the
traceless Hermitian endomorphism of $W^+$, associated with $\Phi$
and $F_A^+$ is the selfdual component of the curvature $F_A$ (with respect to
the Hodge $*$-operator). The section $\Phi$, being a section of $W^+$, is represented
by two forms $(\varphi_0,\varphi_2)$ where $\varphi_0\in\Omega^0(X,E)$,
$\varphi_2\in\Omega^{0.2}(X,E)$.

The Seiberg--Witten equations, as well as the Seiberg--Witten functional $S(A,\Phi)$,
are invariant under the gauge transformations given by the formula
$$
A\longmapsto A+u^{-1}u,\ \Phi\longmapsto u^{-1}\Phi
$$
where $u=e^{i\chi}$ and $\chi$ is a smooth real-valued function so that
$u\in\mathcal G=C^\infty(X,\text{U}(1)$.

Along with Seiberg--Witten equations we shall consider their
perturbed version given by
\begin{equation}
\left\{
\begin{aligned}
\label{eq2}
D_{A}\Phi &=0,\hspace{2cm}\\
F_{A}^{+}+&\eta=\Phi\otimes\Phi^{*}-\frac12|\Phi|^2\cdot\text{Id}
\end{aligned}\right.
\end{equation}
where $\eta$ is a self-dual 2-form on $X$. We shall call this perturbed version
of Seiberg--Witten equations by the $\text{SW}_{\eta}$-equations.
The introduced perturbation is necessary
to guarantee the existence of a solution of Seiberg--Witten equations.

\subsection{Seiberg--Witten equations on a K\:ahler surface}
\label{ssec3.3}

Suppose now that $X$ is a K\"ahler surface, i.e. a smooth compact 2-dimensional complex
manifold. Then the complexified bundle $\Lambda^2_+\otimes\mathbb C$ of selfdual
2-forms on $X$ is decomposed into the direct
sum of subbundles
$$ \Lambda^2_+\otimes\mathbb
C=\Lambda^{2,0}\oplus\mathbb C[\omega]\oplus\Lambda^{0,2}.
$$
Accordingly, the second Seiberg--Witten equation for the curvature
decomposes into the sum of three equations --- one for the
component, parallel to $\omega$, one for $(0,2)$-component and
another one for $(2,0)$-component which is conjugate to $(0,2)$-component and by
this reason is omitted below.

Then the $\text{SW}_{\eta}$-equations take the form
\begin{equation}
\label{eq2}
\left\{
\begin{aligned}
\bar\partial_{B}\varphi_0 &+\bar\partial_{B}^{*}\varphi_2=0,\hspace{3,5cm}\\
F_{B}^{0,2}+&\ \eta^{0,2}=\frac{\bar{\varphi}_0\varphi_2}{2},\hspace{3,4cm}\\
F_{A_{\text{can}}}^{\omega} &+F_{B}^{\omega}=\frac{i}4
(|\varphi_0|^2-|\varphi_2|^2)-\eta^{\omega}.
\end{aligned}\right.
\end{equation}

The first of these equations is the Dirac equation, the second one corresponds to the
(0,2)-component of the curvature equation, and the third one corresponds to the component
of the curvature equation, parallel to $\omega$.

For the $\text{SW}_{\eta}$-equations on a
K\"ahler surface an analogue of Bradlow theorem for vortex equations
on a compact Riemann surface holds. Let $E\to X$ be a Hermitian line
bundle over $(X,\omega,J)$. Suppose that for some $\lambda>0$ its
first Chern class satisfies the inequality
\begin{equation}
0\leq c_1(E)\cdot[\omega]<\frac{c_1(K)\cdot[\omega]}{2}
+\lambda\text{Vol}\,(X)
\end{equation}
where $K$ is the canonical bundle of $X$.
This inequality plays the same role, as the stability condition
$c_1(L)<\tau/4\pi\,\text{Vol}_{g}\,(X)$ in Bradlow theorem.

Under this condition the moduli space of
$\text{SW}_{\eta}$-solutions with the form $\eta=\pi i\lambda\omega$
and $\text{Spin}^c$-structure $W_E$ admits the following
description: \textit{there exists a bijective correspondence between
the gauge classes of $\text{SW}_{\eta}$-solutions
and effective divisors of degree $c_1(E)$ on $X$}.

\subsection{Seiberg--Witten equations on a 4-dimensional symplectic manifold}
\label{ssec3.4}

Suppose now that $X$ is a compact symplectic
4-manifold provided with symplectic 2-form $\omega$ and compatible
almost complex structure $J$. Let $E\to X$ be a Hermitian line
bundle with a Hermitian connection $B$ and
$W_E:=W_{\text{can}}\otimes E$ is the associated spinor bundle.

We take the perturbation form $\eta$ equal to $\eta=-F_{A_{\text{can}}}^{+}+\pi
i\lambda\omega$ with $\lambda>0$.

The corresponding $\text{SW}_{\eta}$-equations have the form
\begin{equation}
\label{sw1}
\left\{
\begin{aligned}
\bar\partial_B\varphi_0 &+\bar\partial_{B}^{*}\varphi_2=0,\\
F_{A_{\text{can}}}^{0,2} &+F_{B}^{0,2}+
\eta^{0,2}=\frac{\bar{\varphi_0}\varphi_2}2,\\
F_{A_{\text{can}}}^{\omega} &+F_{B}^{\omega}+
\eta^{\omega}=\frac{|\varphi_2|^2-|\varphi_0
|^2}{4}
\end{aligned}\right.
\end{equation}
where $(\varphi_0,\varphi_2)\in\Omega^0(X,E)\oplus
\Omega^{0,2}(X,E).$

We introduce now the normalized sections:
$$
\alpha:=\frac{\varphi_0}{\sqrt{\lambda}},\quad
\beta:=\frac{\varphi_2}{\sqrt{\lambda}}.
$$
Then the $\text{SW}_{\eta}$-equations will rewrite as
\begin{equation}
\label{sw2}
\left\{
\begin{aligned}
\bar\partial_{B}\alpha &+\bar\partial_{B}^{*}\beta=0,\\
\frac2{\lambda}F_{B}^{0,2} &=\bar{\alpha}\beta,\\
\frac{4 i}{\lambda}F_{B}^{\omega} &=4\pi+|\beta|^2-|\alpha|^2.
\end{aligned}\right.
\end{equation}

According to Taubes \cite{T}, solutions $(\alpha\equiv\alpha_\lambda,\beta\equiv
\beta_\lambda)$ of the perturbed equations have the following behavior for
$\lambda\to\infty$:
\begin{enumerate}
\item $|\alpha_\lambda|\to 1$ everywhere outside the
set of zeros $\alpha_\lambda^{-1}(0)$;
\item $|\beta_\lambda|\to 0$ everywhere together
with 1st order derivatives.
\end{enumerate}

Denote by $C_\lambda:=\alpha_\lambda^{-1}(0)$ the
zero set of $\alpha_\lambda$. The curves
$C_\lambda$ converge in the sense of currents to
some pseudoholomorphic divisor, i.e. a chain $\sum
d_kC_k$, consisting of connected pseudoholomorphic curves $C_k$
taken with multiplicities $d_k$.

Simultaneously, the original Seiberg--Witten equations reduce to a
family of Ginzburg--Landau vortex equations in the complex planes
normal to the curves $C_k$. These families can be identified with
sections of the $d_k$-vortex bundle over $C_k$ (cf. \cite{Ser}).

Conversely, in order to reconstruct the solution of Seiberg--Witten
equations from the family of vortex solutions in normal planes it
should satisfy a nonlinear equation of $\bar\partial$-type.

Thus, we have for the Seiberg--Witten equations on symplectic
4-manifolds the following correspondence, established by the
adiabatic limit:
$$
\left\{\parbox{2,8cm}{solutions
of Seiberg--Witten equations}\right\}\longmapsto
\left\{\parbox{5cm}{families of vortex solutions in normal planes of
pseudoholomorphic divisors}\right\}.
$$


\begin{thebibliography}{JT999}

\bibitem{B}
S.B.Bradlow, \textit{Vortices in holomorphic line bundles over
closed K\"{a}hler manifolds}, Comm. Math. Phys. \textbf{135}(1990),
1--17.

\bibitem{JT}
A.Jaffe, C.H.Taubes, \textit{Vortices and monopoles}.-Birkh\"{a}user:
Boston, 1980.

\bibitem{LP}
E.M.Lifshitz, L.P.Pitaevskii, \textit{Statistical Physics, part 2.
Theory of the condensed state}, Pergamon Press: London, 1980.

\bibitem{LM}
H.B.Lawson Jr., M-L.Michelson, \textit{Spin geometry}.-Princeton:
Princeton University Press, 1989.

\bibitem{Ma}
N.S.Manton, \textit{A remark on the scattering of  BPS monopoles},
Phys. Lett. \textbf{110B}(1982), 54--56.

\bibitem{Pal}
R.V.Palvelev, \textit{Justification of the adiabatic principle in
the Abelian Higgs model}, Trans. Moscow Math. Soc., \textbf{72}
(2011), 219--244.

\bibitem{Pal-Ser}
R.V.Palvelev, A.G.Sergeev, \textit{Justification of the adiabatic
principle for hyperbolic Ginzburg-Landau equations}, Proc. Steklov
Inst. Math. \textbf{277}(2012), 191--205.

\bibitem{SW}
N.Seiberg, E.Witten, \textit{Monopoles, duality and chiral symmetry
breaking in $N=2$ supersymmetric Yang-Mills theory}, Nucl.Phys.
\textbf{B426}(1994), 581--640.

\bibitem{Ser}
A.G.Sergeev, \textit{Adiabatic Limit in the Ginzburg-Landau and
Seiberg-Witten equations}, Proc. Steklov Inst. Math., 289 (2015),
227--285

\bibitem{T}
C.H.Taubes, $SW \Rightarrow Gr$: \textit{From the
Seiberg-Witten equations to pseudo-holomorphic curves},
J. Amer. Math. Soc. \textbf{9}(1996), 845--918.

\end{thebibliography}
\end{document}